\documentclass[a4paper,12pt]{article}

\usepackage{epsfig}

\usepackage{amsmath}
\usepackage{amssymb}
\usepackage{cite}

\usepackage{fancyhdr}
\newcommand{\chf}{\,{_1{\cal F}_1}} 
\newcommand{\bo}[1]{{\cal O}\big(#1\big)}

\newcommand{\beqs}{\begin{equation*}}
\newcommand{\beq}{\begin{equation}}

\newcommand{\eeqs}{\end{equation*}}
\newcommand{\eeq}{\end{equation}}

\newcommand{\beqas}{\begin{eqnarray*}}
\newcommand{\beqa}{\begin{eqnarray}}

\newcommand{\eeqas}{\end{eqnarray*}}
\newcommand{\eeqa}{\end{eqnarray}}

\newcommand{\eq}[2]{\begin{equation} #1 \label{#2} \end{equation}}

\newcommand{\eps}{\varepsilon}
\newcommand{\al}{\alpha}

\newcommand{\om}{\omega}

\newcommand{\la}{\lambda}

\newcommand{\blist}{\begin{itemize}}

\newcommand{\elist}{\end{itemize}}

\providecommand{\href}[2]{#2}

\newcommand{\twod}{$2D$}

\DeclareFontFamily{OT1}{rsfs}{}
\DeclareFontShape{OT1}{rsfs}{m}{n}{ <-7> rsfs5 <7-10> rsfs7 <10->rsfs10}{} 
\DeclareMathAlphabet{\mycal}{OT1}{rsfs}{m}{n}

\providecommand{\href}[2]{#2}

\newcommand{\rnKH}{rnKH}

\begin{document}

\begin{titlepage}

\renewcommand{\thefootnote}{\fnsymbol{footnote}}

\hfill TUW--03--22

\begin{center}
\vspace{0.5cm}

{\Large\bf On static solutions in 2D dilaton gravity with scalar matter}

\vspace{1.0cm}

{\bf D.\ Grumiller\footnotemark[1] and D.\ Mayerhofer\footnotemark[2]
}
\vspace{7ex}

  {\footnotemark[1] \footnotemark[2]
\footnotesize Institut f\"ur
    Theoretische Physik, Technische Universit\"at Wien \\ Wiedner
    Hauptstr.  8--10, A-1040 Wien, Austria}
\vspace{2ex}

  {\footnotemark[1]\footnotesize Institut f\"{u}r Theoretische Physik, Universit\"{a}t Leipzig,\\Augustusplatz 10--11, D-04109 Leipzig, Germany}

   \footnotetext[1]{e-mail: \texttt{grumil@hep.itp.tuwien.ac.at}}
   \footnotetext[2]{e-mail: \texttt{mayerhofer@teilchen.at}}

\end{center}
\vspace{7ex}

\begin{abstract}

Within the first order formalism static solutions of generic dilaton gravity in 2D with self-interacting (scalar) matter can be discussed with ease. The question of (non)existence of Killing horizons is addressed and the interplay with asymptotic conditions is investigated. Naturally, such an analysis has to be a global one. A central element in the discussion is the rank of the Jacobi matrix of the underlying dynamical system. With some (pathological) exceptions Killing horizons exist only if it equals to 3. For certain self-interactions asymptotically flat black holes with scalar hair do exist. An example relevant to general relativity is provided. Finally, generalizations are addressed including 2D type 0A string theory as a particular example. Additionally, in a pedagogical appendix the mass definition in dilaton gravity is briefly reviewed and a unique prescription to fix scaling and shift ambiguity is presented.

\end{abstract}

\vfill
\end{titlepage}

\section{Introduction}

Spherically symmetric static solutions of general relativity with a scalar matter field have a long history in physics \cite{Fisher:1948yn} and have been rediscovered and discussed many times \cite{recentscalar} in slightly varying contexts. In particular, in \cite{Bronnikov:2001ah} several theorems have been proven regarding the absence of hairy black hole (BH) solutions (cf.~\cite{olderlit} for some earlier literature).

If no Killing horizon exists static solutions with scalar matter are capable to violate the cosmic censorship hypothesis---this is true for the Fisher solution \cite{Fisher:1948yn}, but also for related nonstatic solutions like the self-similar one given by Roberts \cite{Roberts:1989sk}. 
On the other hand, if a Killing horizon exists and the solution is not Schwarzschild then the no-hair conjecture is violated. A discussion of hairy BHs as bound states between BHs and solitons can be found in \cite{Ashtekar:2000nx}.

It is well known \cite{dilaton} that spherical symmetry reduces the original model effectively to a twodimensional ($2D$) one, namely a specific $2D$ dilaton gravity theory. Generic dilaton gravity was found to be of interest on its own (e.g.~in studies of BH evaporation), and it has overlaps with string theory, integrable models and noncommutative geometry (for a review on dilaton gravity cf.~\cite{Grumiller:2002nm}).

Consequently, it is not only natural but also of interest to inquire about the existence of Killing horizons in static solutions of generic dilaton gravity with scalar matter \cite{Filippov:2002sp}. It should be pointed out that the analysis in that reference is a local one in the vicinity of the Killing horizon. Thus, the question whether an {\em isolated static BH (immersed in flat spacetime) with scalar hair} may exist cannot be addressed. By ''immersed in flat spacetime'' we mean that a one parameter family of solutions exists (labelled by a constant of motion which shall be called ''mass'' and be denoted by $M$) such that for a certain value of the parameter (typically $M=0$) flat spacetime is a solution of the corresponding dilaton gravity model. It will be made explicit below that, indeed, such a constant of motion always exists and thus this notion is meaningful.

The main goal of the present work is to provide an answer to the question formulated in the last paragraph in the context of generic dilaton gravity with scalar matter: it will be negative in the absence of self-interactions, which is physically plausible because if gravity were the only force due to its exclusively attractive nature it cannot support hairy BHs. It will be affirmative in the presence of self-interactions, provided the obstructions on the potential derived below can be circumvented. The analysis has to be a global one because one has to connect the asymptotic behavior with the one close to eventual Killing horizons. In this context the definition of the mass of a BH is of relevance, which is why a brief review on this issue is presented. A unique prescription fixing the shift and scale ambiguity is given valid for generic dilaton gravity in $2D$.

Besides the obvious application to general relativity with a self-interacting scalar matter field also an example relevant to string theory is discussed in detail, namely $2D$ type 0A with a static tachyon. Moreover, spherically symmetric dilaton BHs in $4D$ with static quintessence field fit into our general scheme.

This paper is organized as follows: Section \ref{se:1} briefly recapitulates dilaton gravity in first order formulation. In section \ref{se:eom} the relevant equations of motion for static solutions of generic dilaton gravity with scalar matter are presented and simplified to a system of three coupled ordinary differential equations (ODEs) which are of first order in geometry and of second order in the matter degrees of freedom. Section \ref{se:2} and \ref{sec:SKH} are devoted to the question of existence of Killing horizons in the presence of constraints on the asymptotic behavior of geometry and matter. For sake of definiteness the focus will be mainly on spherically reduced gravity. Section \ref{se:2} provides several nogo results while section \ref{sec:SKH} discusses those cases circumventing the former. Possible generalizations are addressed in section \ref{se:3} and $2D$ type 0A string theory with static tachyon is treated as an application. The appendix contains the proper mass definition in generic dilaton gravity and provides several examples, thereby resolving some slight puzzlement that arose in the context of $2D$ type 0A string theory.

\section{Recapitulation of dilaton gravity}\label{se:1}

The purpose of this section is to fix our notation. For details on dilaton gravity the review ref.\ \cite{Grumiller:2002nm} may be consulted. In its first order version the generalized dilaton gravity (GDT) action reads
\eq{
L=L^{(FOG)}+L^{(m)}\,,
}{vbh:action} 
with the geometric part
\eq{
L^{(FOG)} = \int\limits_{\mathcal{M}_2} \, 
\left[ X_a \left(D\wedge e\right)^a + X d\wedge\omega + \epsilon \mathcal{V} 
(X^a X_a, X) \right]\,.
}{vbh:geometryaction}
The notation of ref.\ \cite{Grumiller:2002nm} is used: $e^a$ is the
zweibein one-form, $\epsilon = e^+\wedge e^-$ is the volume two-form. The one-form
$\omega$ represents the  spin-connection $\om^a{}_b=\eps^a{}_b\om$
with  the totally antisymmetric Levi-Civit{\'a} symbol $\eps_{ab}$ ($\eps_{01}=+1$). With the
flat metric $\eta_{ab}$ in light-cone coordinates
($\eta_{+-}=1=\eta_{-+}$, $\eta_{++}=0=\eta_{--}$) the first
(``torsion'') term of (\ref{vbh:geometryaction}) is given by $X_a(D\wedge e)^a =
\eta_{ab}X^b(D\wedge e)^a =X^+(d-\omega)\wedge e^- +
X^-(d+\omega)\wedge e^+$. Signs and factors of the Hodge-$\ast$ operation are defined by $\ast\epsilon=1$. The fields $X,X^a$ can be interpreted as Lagrange multipliers for curvature and torsion, respectively. They enter the potential $\mathcal{V}$ which is assumed to have the standard form
\begin{equation}
\mathcal{V}(X^a X_a, X) = X^+X^- U(X) +V(X)\,. \label{vbh:calV}
\end{equation}
If not stated otherwise we will restrict ourselves to the relevant special case $U=-a/X$. The action (\ref{vbh:geometryaction}) is classically
equivalent \cite{Kummer:1998zs} to the more familiar general second order 
dilaton action\footnote{Obviously this equivalence implies that after elimination of auxiliary fields the equations of motion are identical for both formalisms. However, already at the classical level it is usually much simpler to employ the first order formulation. The most recent demonstration of this fact is contained in refs.~\cite{Guralnik:2003we,Grumiller:2003ad}: in \cite{Guralnik:2003we} the second order formulation has been employed and by ingenuity a non-trivial solution has been found locally, while in \cite{Grumiller:2003ad} the application of the first order formulation allowed a straightforward global discussion of all classical solutions. In the current paper the advantages of the first order formulation are twofold: first of all, it is very simple to obtain the ``reduced set of equations of motion'' (\ref{eq:hamilton})-(\ref{eq:kleingordon}) below; second, the discussion of the ``mass'' becomes particularly transparent due to its relation to the Casimir function in the context of Poisson-$\sigma$ models (PSMs) --- cf.~appendix \ref{app:mass}.}
\begin{equation}
 L^{(SOG)}=\int\limits_{\mathcal{M}_2} \, d^2 x \sqrt{-g} \left[
X\frac R2 -\frac{U(X)}2 (\nabla X)^2 +V(X) \right]\,. 
\label{vbh:sog}
\end{equation}
For the matter part we choose the action of a (non-minimally coupled) scalar field $\phi$ 
\eq{
L^{(m)} =  \int\limits_{\mathcal{M}_2} \, F(X) \left(\frac{1}{2} d\phi \wedge 
\ast d\phi + \epsilon f(\phi)\right)\,,
}{vbh:matteraction}
with an---in principle arbitrary---coupling function $F(X)$. In practice
the cases $F=\rm const.$ (minimal coupling) and $F\propto X$ are the most
relevant ones. The self-coupling function $f(\phi)$ will play a rather important role for the (non)existence of regular, nontrivial Killing horizons.

\section{Equations of motion in static limit}\label{se:eom}

For convenience Sachs-Bondi gauge
\eq{
e_1^+=0\,,\quad e_1^-=-1\,,\quad e_0^+=(e)\,,\quad e_0^- = \frac h2\,,
}{eq:SB}
will be employed. $(e)$ and $h$ are arbitrary functions of the worldsheet coordinates. Additionally, the dilaton can be fixed as the ``radial coordinate''\footnote{This is possible as long as $dX\neq 0$. If $dX=0$ in an open region a ``constant dilaton vacuum'' emerges for that region; these solutions are essentially trivial and therefore of limited interest (cf.~footnote \ref{fn:cdv}). If $dX=0$ on an isolated point it is called a bifurcation point (in the language of general relativity it corresponds to the ''bifurcation 2-sphere'' encountered in the Schwarzschild solution). By imposing Sachs-Bondi gauge (\ref{eq:SB}) it is not possible to cover such points (this is well-known for the Schwarzschild BH and it remains true for the general case: in the first order formalism employed in this work the technical assumption $X^+\neq 0$ (or alternatively $X^-\neq 0$) -- a crucial ingredient to derive the line element in Sachs-Bondi gauge -- breaks down whenever $dX=0$ because at such  points $X^+=0=X^-$; cf.~the ``first obstruction'' in Section~\ref{sec:3.1} below). Thus the fixing in (\ref{eq:dilfix}) does not impose an additional restriction. If one is interested in the behavior around the bifurcation point the methods of \cite{Walker:1970,Klosch:1996qv} may be applied. As the main goal of this work is to establish criteria for (non-)existence of asymptotically flat BH solutions with scalar hair these subtleties are of no relevance here.}
\eq{
X=b r^\al\,,\quad b,\al >0\,.
}{eq:dilfix}
In general $b$ is dimensionful and $\al$ dimensionless.
As coordinates we choose $x^0=v$ and $x^1=r$. Furthermore the abbreviations $\partial_0 (\bullet) = \partial/\partial x^0 (\bullet)$ and $\partial_1 (\bullet) = \partial/\partial x^1 (\bullet) = (\bullet)'$ are used, where $\bullet$ stands for $(e),h,\omega_\mu,X^\pm,X$ and $\phi$. By staticity we mean
\eq{
\partial_0 (\bullet) = 0\ .
}{eq:stat}
Thus, all fields have to depend on $r$ only. Actually, it is sufficient to impose this condition solely on the geometric fields, $(e), h, \om_\mu, X^\pm, X$, because the conservation law, equation \eqref{eq:sn10} below, then implies $\partial_0\phi=0$.\footnote{At first glance a second solution with time-dependent $\phi$, fulfilling $\partial_0\phi+h(x^1)\partial_1\phi=0$, appears to be possible. However, a careful analysis of all equations of motion reveals that this solution allows only for constant $f(\phi)$, a case which is of little interest as a constant term in $f$ can always be eliminated by redefining $V$.}

The line element for Sachs-Bondi gauge now reads
\eq{
(ds)^2=(e)\left(hdv-2dr\right)dv\,.
}{eq:line}
The quantities $(e)$ and $(e)h$ can be interpreted, respectively, as (square root of minus the) determinant of the metric and as Killing norm. Thus, for regular $(e)$, zeros of $h$ correspond to Killing horizons.

The variation of \eqref{vbh:geometryaction} with respect to $\omega$ fixes the auxiliary fields to $X^+ = -X'$ and $X^- = -X'\frac{h}{2(e)}$. Furthermore the variation $\delta X^\pm$ provides
\begin{equation}
\begin{aligned}
        \om_0 &= \frac h2 \left[\ln'{(h/2)}+\ln'{(e)}+\frac{2 a \al}{r}\right]\ ,\\
        \om_1 &= -\ln'{(e)}-\frac{a \al}{r}\ .\\
\end{aligned}
\end{equation}
Certain combinations of $\delta e^\mp$ and $\delta\phi$ can be used to simplify the equations of motion (e.o.m.) to a system of three nonlinear coupled ODEs. It is convenient to set $1-\al=-a\al$ (unless $a=1$). Moreover, the redefinition
\eq{
Z:=-(X^+e_0^-+X^-e_0^+)=b\al r^{\al-1}h
}{eq:Z}
is helpful. With these simplifications the three ODEs read
\begin{eqnarray}
&& \ln'{(e)} + \frac{2}{b\al r^{\al-1}}F(br^\al)(\phi')^2= 0\,, \label{eq:hamilton} \\
&& Z' + 2(e)\left[V(br^\al)+F(br^\al)f(\phi)\right] = 0\,, \label{eq:slicing} \\
&& \left(\frac{F(br^\al)}{b\al r^{\al-1}}Z\phi'\right)' = (e)F(br^\al)\frac{\partial f}{\partial \phi}\,. \label{eq:kleingordon}
\end{eqnarray}
These equations are the dilaton gravity generalization (with restrictions on $U$ and $X$ as mentioned above) of the standard e.o.m. found in topically similar literature and can be interpreted, respectively, as Hamilton constraint, slicing condition\footnote{This name has been chosen in accordance with literature on spherically symmetric collapse \cite{Choptuik:1993jv,Gundlach:1998wm}. Note, however, that $f(\phi)$ enters this equation, so it is not a purely geometrical one in general.} and Klein-Gordon equation (with potential). 

From a dynamical point of view eqs.\ (\ref{eq:hamilton})-(\ref{eq:kleingordon}) build a non-autonomous\footnote{The ``time'' is nothing but the radius $r$ or the dilaton field $X$.} dynamical system with phase space coordinates $z_i=(q_1,q_2,p_1,p_2)$ where $q_1=\phi$, $q_2=\ln{(e)}$, $p_1=\pi=\phi'ZF/(b\al r^{\al-1})$, $p_2=Z$. Thus, an evolution equation $z_i'=A_i(z,r)$ is encountered with a given vector field $A_i$, 
\eq{
A_{1,2} = \frac{b\al r^{\al-1}}{F(br^\al)}\frac{\pi}{Z}\begin{pmatrix}1 \\ -2\frac{\pi}{Z}\end{pmatrix}\,,\quad A_{3,4} = (e)F(br^\al)\begin{pmatrix}\frac{\partial f}{\partial \phi}\\-2\left(f+\frac{V(br^\al)}{F(br^\al)}\right)\end{pmatrix}\,.
}{eq:stat8}
Singular points $A_i(z,r_0)=0$ exist only if $\pi r^{\al-1}/(ZF)=0$ at $r=r_0$. In addition, either $(e)F$ must vanish there, or $\partial f/\partial \phi$ must vanish and $f$ must be tuned to fulfill $f=-V/F$ at the same point.

Relevant for dynamical properties is the Jacobian $M_{ij}=\partial A_i/\partial z_j$. Obviously the relations
\eq{
\frac{\partial A_{1,2}}{\partial q_i}=0\,,\quad\frac{\partial A_{3,4}}{\partial p_i}=0\,,
}{eq:stat9}
considerably simplify calculations, because $M_{ij}$ acquires block
form with two null matrices in the block diagonal. Thus, its rank
is just the sum of the ranks of the two off-diagonal $2\times 2$ submatrices. Let us first consider $\partial A_{1,2}/\partial p_i$: since only the combination $\pi/Z$ appears it cannot have full rank. On the other hand, its rank cannot vanish identically unless $\pi\equiv 0$, which we rule out because we are interested in nontrivial solutions only ($\pi=0$ for $F,Z\neq 0$ implies $\phi'=0$ and thus matter would be absent). Therefore, the rank of this submatrix for nontrivial configurations always equals to 1. Hence, the rank of the full matrix crucially depends on the other submatrix $\partial A_{3,4}/\partial q_i$. There are three possibilities: rank 2, 1 or 0, implying for the full Jacobi matrix rank 3, 2 or 1, respectively. Thus, at least one constant of motion must exist which can be identified with the ADM mass in those cases where this notion makes sense (otherwise it is related to a conserved quantity which exists generically in {\twod} dilaton gravity); for simplicity this quantity will be called ``mass''. A proper way to fix the scaling and shift ambiguity inherent in any mass definition is presented in appendix \ref{app:mass}. Assuming that $(e)\neq 0\neq F$ for regularity this establishes simple criteria for classification.

\paragraph{Jacobi matrix with rank 1}
By inspection of the right equation in (\ref{eq:stat8}) this is possible if and only if $f=\rm const.$ and $V=-Ff$. However, the constant part of $f$ can be absorbed always into the geometric potential $V$ by a redefinition of the latter and thus without loss of generality this case equals to $f\equiv 0 \equiv V$. The three constants of motion are mass, $\pi$ and $Z$. Matter and also geometry are trivial.

\paragraph{Jacobi matrix with rank 2}
The scalar potential has to fulfill the differential equation 
\begin{equation}
\label{eq:rank2DE}
	f''(\phi)\ (f(\phi)+\tfrac{V}{F}) = (f'(\phi))^2 .
\end{equation}
Here the prime denotes the derivative with respect to the argument. When no argument is specified prime stands for the derivative with respect to $r$. There are two simple possibilities to fulfill \eqref{eq:rank2DE}: either $f\equiv 0$ and $V\neq 0$; in this case the two constants of motion are mass and $\pi$ (e.g.\ the Fisher solution \cite{Fisher:1948yn}). Or $f\propto f'$ and $V\equiv 0$; then the constants of motion are mass and $\pi+aZ$, where $a\in\mathbb{R}$ (e.g.\ the polarized Gowdy model \cite{Gowdy:1971jh}). For the general case the second constant of motion can only be determined up to an integral
\begin{equation}
\label{eq:rank2COM}
	\pi + \int\!\!dr Z' \frac{f''(\phi)}{f'(\phi)} = c .
\end{equation}

\paragraph{Jacobi matrix with rank 3}
This is the generic case for nontrivial $f$, where ''nontrivial'' refers to the fact that $f$ must not fulfill \eqref{eq:rank2DE}. The only constant of motion that remains is the mass. As it will turn out only for this case regular, nontrivial Killing horizons may emerge (with a few
somewhat pathological exceptions).

\section{No Killing horizons}\label{se:2}

\subsection{Definition of a regular nontrivial Killing horizon}

We require the existence of a particular (finite) value of the dilaton, $X=X_h$ at which the following conditions are fulfilled: 
\eq{
h=0=Z\,,\quad |(e)|>0\,,\quad |\bullet|<\infty\,, 
}{eq:stat6}
where $\bullet$ stands for $X^+,X^-,\om_0,\om_1,(e),h$ and $\phi$ as well as for their first derivatives. In particular, also $\phi'$ has to remain finite at a Killing horizon.

Additionally (this is what ``nontrivial'' refers to) $\phi$ must not be constant globally---otherwise the system essentially reduces to a matterless dilaton gravity theory, the classical solutions of which are well-known \cite{Klosch:1996fi,Klosch:1996qv}. The Hamiltonian constraint then implies that also $(e)$ cannot be constant. 

In the rest of this section several no-go results regarding the existence of such Killing horizons will be provided. We will abbreviate the term ``regular nontrivial Killing horizon'' by \rnKH.

\subsection{Absence of self-interactions implies absence of \rnKH}

In the absence of a nontrivial external potential (i.e.\ for $f(\phi)=\rm const.$) eq.~\eqref{eq:kleingordon} can be integrated immediately\footnote{This scenario corresponds either to the rank=1 case (for $V\equiv 0)$ or to one of the rank 2 cases (for generic $V$). One can reduce the whole system to a single nonlinear second order ODE. For SRG the Fisher solution can be reproduced \cite{Fisher:1948yn}. As noted before one constant of motion turns out to be the ADM mass while the other one is the ``family parameter''.}
\eq{ 
\frac{F(br^\al)}{b\al r^{\al-1}}Z\phi' = c\,,\quad c\in\mathbb{R}
}{eq:stat1}
This implies that for nonvanishing $c$ no {\rnKH} can exist: if $Z=0$ at some finite $r$ then $\phi'$ must diverge. For $c=0$, however, $Z\phi'$ has to vanish everywhere. The simplest cases are either $Z\equiv 0$ or $\phi'\equiv 0$. 
The former case is trivial ($r$ has to be reinterpreted as a light-like coordinate), while the latter one implies $(e)=\rm const.$ by virtue of (\ref{eq:hamilton}). Then (\ref{eq:slicing}) becomes a linear ODE which can be solved trivially. By choosing $V$ accordingly any number of Killing horizons is possible. However, this case cannot be considered as nontrivial, because the condition $\phi'\equiv 0$ just implies that, in fact, no matter degrees of freedom are present. Therefore, the discussion reduces to dilaton gravity without matter. But this possibility has been ruled out in our definition of \rnKH.\footnote{If $Z\equiv 0$ or $\phi'\equiv 0$ is not validy globally but only patchwise the same conclusion holds for the reasons discussed above; the only nontrivial additional considerations concern the hypersurface of patching: to guarantee the absence of induced localized matter fluxes $Z, (e)$ and $Z'$ have to be continuous. Thus, the hypersurface of patching has to be an extremal Killing horizon, which can be achieved in the matterless regions by tuning the geometric potential $V$ accordingly. The simplest example is extremal Reissner-Nordstr\"om patched to a $Z\equiv 0$ region at the Killing horizon $R=R_0$: $(ds)^2=2dudR+\theta(R_0-R)(1-R_0/R)^2(du)^2$ with $dR=dr((e)\theta(r-r_0)+\theta(r_0-r))$ and $r_0=R_0$. Note that the presence of ``matter'' in the $Z\equiv 0$ region $R>R_0$ is irrelevant as it only modifies the relation between $R$ and $r$ as seen from (\ref{eq:hamilton}).}

It is straightforward to generalize this discussion to arbitrary potentials $U(X)$ in (\ref{vbh:calV}). The only difference is that for $c=0$ the determinant $(e)$ no longer is constant but a certain function of $r$. 

In conclusion, no {\rnKH} exists for arbitrary static configurations of generalized dilaton gravity theories in {\twod} of type (\ref{vbh:action})-(\ref{vbh:matteraction}) if $f(\phi)\equiv 0$.

\subsection{Rank $\leq 2$ of Jacobi matrix implies absence of rnKH}

If $f\equiv 0$ the discussion in the previous subsection can be applied. According to the analysis at the end of section \ref{se:eom} another simple possibility is $V\equiv 0$ and $f(\phi)\propto \exp{(k\phi)}$, $k\in\mathbb{R}$, implying rank 2. This is reflected by the fact that (\ref{eq:kleingordon}) together with (\ref{eq:slicing}) allow for a first integral
\eq{
\frac{F(br^\al)}{b\al r^{\al-1}}Z\phi' + \frac k2 Z = c\,,\quad c\in\mathbb{R}\,.
}{eq:stat2}
For nonvanishing $c$ the Killing horizon condition $Z=0$ implies a diverging $\phi'$ and thus no {\rnKH} exists. For $c=0$ either $Z$ has to vanish everywhere or $F \phi' \propto r^{\al-1}$ everywhere. The first alternative is trivial, the second one turns out to be pathological for $F\propto X$: as $\phi\propto\ln{r}$ the scalar field diverges in the ``asymptotic region'', which may be located either at $r=0$ or at $r=\infty$. Thus, no regular BH solution emerges in such a way.  

For the general rank 2 case, with $f(\phi)$ being a general solution of \eqref{eq:rank2DE}, the constant of motion \eqref{eq:rank2COM} can be integrated perturbatively from $r$ to $r+\epsilon$ (with $\epsilon\ll 1$) yielding a linear dependence on $Z$:
\begin{equation}
	 Z\left(\frac{F(br^\al)}{b\al r^{\al-1}} \phi' + \frac{1}{2} \frac{f''(\phi)}{f'(\phi)} - \frac{1}{2}\ \varepsilon\,  \partial_r\!\left(\frac{f''(\phi)}{f'(\phi)}\right)\right) + \bo{\varepsilon^2} = c
\end{equation}
Similar to the above discussion for a nonvanishing constant $c$ the Killing horizon condition implies that at least one of the terms in the bracket has to diverge, which again is in conflict with our definition of a rnKH. For $c=0$, if $Z$ vanishes identically again, a trivial solution is encountered. Otherwise, suppressing all terms containing $\epsilon$, the remaining PDE can be integrated easily:
\begin{equation}
\label{eq:r2int}
	\int\!\! \frac{d\phi}{\frac{\partial}{\partial \phi} \ln \left(\frac{\partial}{\partial \phi} f(\phi)\right)} = - \frac{b \alpha}{2} \int\!\! dr \frac{r^{\alpha-1}}{F(b r^\alpha)} + \tilde{c}
\end{equation}
With $F(X) = - \gamma X^\delta$ the right hand side is proportional to
\begin{equation}
	\begin{aligned}
		\ln r & \quad & \text{for} & \quad & \delta = 1 \\
		r^{\alpha(1-\delta)} & \quad & \text{for} & \quad & \delta \neq 1
	\end{aligned}
\end{equation}
If $f(\phi)$ is expressed by a power series\footnote{The terms $a_0$ and $a_1$ are excluded because a constant term can always be transfered to the geometric potential $V(r)$, and a term linear in $\phi$ can be absorbed by a redefinition of $\phi$. The special case \eqref{eq:stat2} presents an example, where no term dominates asymptotically.} $f(\phi) = \sum_{k\neq 0,1} a_k \phi^k$ and one term ($k=m$) dominates asymptotically, the integration of the left hand side of \eqref{eq:r2int} gives $\tfrac{1}{2}(m-1)\phi^2$. Therefore we can conclude that for $\delta=1$, as it is the case for spherically reduced gravity, $\phi^2$ is proportional to $\ln r$ and hence diverges for $r \rightarrow \infty$ and $r \rightarrow 0$. Considering $\delta \neq 1$ $\phi$ diverges for $\delta < 1$ and $r \rightarrow \infty$ or $\delta > 1$ and $r \rightarrow 0$. As we require asymptotic flatness, the cases of diverging $\phi$ also implicate the nonexistence of a rnKH.

Thus, also for the rank 2 scenario no {\rnKH} can exist with the (typically pathological) exceptions discussed above. It should be noted that the latter arise only if one of the constants of motion is infinitely finetuned---thus, if open regions of ``initial data'' are considered they essentially disappear.

\subsection{Generic obstructions from positivity properties}

As in many other branches of physics features of positivity and convexity simplify the discussion of the global behavior of the solutions and provide the basis for eventual obstructions.

The first remark concerns the quantity $(e)$: if it vanishes at a certain point then the metric (\ref{eq:line}) degenerates. Thus one can require, say, positivity of $(e)$ in a regular patch. This condition will be imposed for the rest of this work.

If $F(X)$ has a definite sign then from the Hamiltonian constraint one can deduce that also $(e)'$ has a definite sign. For physical reasons $F$ should be negative for any realistic model and thus $(e)$ must be a monotonically increasing function of $r$. Also this condition will be assumed henceforth. In fact, it will be supposed from now on that either $F$ is constant or linear in $X$, because this covers practically all cases discussed in the literature.

If additionally $f$ has a definite sign and $V$ has the same sign
as $f\cdot F$ then also $Z'$ has a definite sign. This implies
immediately that at most one Killing horizon can exist. Moreover,
outside the horizon also $Z$ must have a definite sign. Thus, if
$Z'$ has a definite sign outside the Killing horizon it has to be
positive. After multiplication with $\phi'/((e)F)$ one can
integrate formally\footnote{Following from the general Klein-Gordon
equation \eqref{eq:kleingordon}, equation~\eqref{eq:stat7} is the dilaton gravity generalization (with restrictions on $U$ and $X$ as mentioned above) of the integration used in standard literature like e.g. reference 2 in \cite{olderlit}.} the Klein-Gordon equation from the Killing horizon to the asymptotic region
\eq{
f(\phi)|_{r_\infty}-f(\phi)|_{r_h} = \int\limits_{r_h}^{r_\infty}\left((\phi')^2\left(\frac{FZ}{b\al r^{\al-1}}\right)'\frac{1}{(e)F} + ((\phi')^2)' \frac{Z}{2(e)b\al r^{\al-1}}\right)\,.
}{eq:stat7}
The left hand side obviously has a definite sign. It is given by $-f|_{r_h}$ if asymptotic conditions require that $f(\phi)$ vanishes at $r=r_\infty$. On the right hand side the first integrand has a definite sign according to the previous discussion. If the second term has the same one\footnote{The behavior of the sign of $((\phi')^2)'$ can be studied easily in the asymptotic region by virtue of (\ref{eq:hamilton}), taking into account $\ln{(e)}<0$, $\ln'{(e)}>0$ and (asymptotically) $\ln''{(e)}<0$ for $F<0$, which is the case for spherically reduced gravity. For $F>0$ the signs of $\ln'{(e)}$ and $\ln''{(e)}$ change.}, then also the right hand side of (\ref{eq:stat7}) has a definite sign. This provides an obstruction on the existence of Killing horizons if the sign of the left hand side does not match with the one on the right hand side. For instance, the famous no-hair theorem emerges as a special case of these simple considerations (cf.\ theorem 4 and eqs.\ (20) and (21) of \cite{Bronnikov:2001ah}).

However, it is clear that these obstructions will be relevant only for a certain class of models. Thus, dilaton gravity coupled to selfinteracting scalar matter is capable to circumvent the no-hair theorem.

\section{Killing horizons}
\label{sec:SKH}

To summarize the consequences of the previous section, models which exhibit static solutions with at least one {\rnKH} must have rank 3 of the Jacobian\footnote{The only exception required an infinite finetuning of one of the constants of motion and implied for $F\propto X$ asymptotic non-flatness. Cf.~the discussion below (\ref{eq:stat2}).} derived from (\ref{eq:stat8}) and a nontrivial selfinteraction potential $f(\phi)$. Also, positivity obstructions discussed around (\ref{eq:stat7}) have to be circumvented.

\subsection{Asymptotic behavior}

For models which allow asymptotically flat solutions in the absence of matter it is natural to impose boundary conditions in the asymptotic region $r=\infty$:
\eq{
\phi = 0\,,\quad (e) = 1\,,\quad Z \propto r^{\al-1}
}{eq:stat5}
The way in which $\phi'$ and $f(\phi)$ have to vanish (as well as the next-to-leading order behavior of $(e)$) can be deduced from the equations of motion---the behaviour of $f(\phi)$ is determined by (\ref{eq:slicing}), the behaviour of $\phi'$ then follows from (\ref{eq:kleingordon}) and implies the next-to-leading order scaling of $(e)$ by virtue of (\ref{eq:hamilton}).
Such considerations are relevant if one is interested in asymptotically flat hairy BH solutions.

\subsection{Near horizon approximation and extremality}

Close to a Killing horizon one can expand in powers of $Z$. A particular consequence of such a perturbative treatment is
\eq{
Z'=-2(e)\left[V(br^\al)+F(br^\al)f(\phi)\right]+{\cal O}(Z^2)\,.
}{eq:stat3}

From (\ref{eq:stat3}) one can deduce immediately the condition for the existence of an extremal Killing horizon: $V,F$ and $f$ must be tuned such that at $Z=0$ the condition $V(br^\al)+F(br^\al)f(\phi)=0$ holds. Additionally $df(\phi)/d\phi$ must vanish at this point. These conditions need not be accessible for a given model: together with the Hamilton constraint and eventual positivity properties obstructions on the existence of extremal Killing horizons may be derived in certain cases. Analogous conditions can be derived for higher order Killing horizons.

%
%
%
%
%

\subsection{Specialization to spherically reduced gravity}
\label{sec:horSRG}

In this section the previous general results are specialized to spherically reduced gravity ($V(X) = -b$, $F(X)=-\gamma X$ with $\gamma > 0$, $X=b r^\alpha$ and $\alpha=2$ resp. $a=1/2$) and one example is worked out explicitly. Results of this section are based upon \cite{may04}; the simpler case $f=0$ has been studied in the same formalism in appendix D.6 of \cite{Grumiller:2001ea}.

The equations of motion (\ref{eq:hamilton} - \ref{eq:kleingordon}) can be rewritten as
\begin{equation}
        \label{eq:SRGHC}
        \ln'(e) = \gamma\ r\ (\phi')^2\,,
\end{equation}
\begin{equation}
        \label{eq:SRGSC}
        \tilde{Z}' = (e)\left[1 + \gamma\ r^2 f(\phi) \right]\,, \qquad \tilde{Z}' = h'r+h\,,
\end{equation}
\begin{equation}
        \label{eq:SRGKG}
        \left(r \tilde{Z} \phi'\right)' = (e)\, r^2 \frac{\partial f}{\partial \phi}\,,
\end{equation}
and are found in a similar form in standard literature for spherically reduced gravity \cite{olderlit}. Here $Z$ was redefined to $Z = 2 b \tilde{Z}$. As mentioned above regularity of the metric is demanded and therefore $(e) \neq 0$ for all radii. So the condition of asymptotic flatness ($(e) \rightarrow 1$)  together with \eqref{eq:SRGHC} lead to $(e)<1$ and $(e)'>0$. An examination of \eqref{eq:SRGSC} shows that one can distinguish two regions where the sign of $\tilde{Z}'$ restricts the possible values of $f(\phi(r))$:
\begin{equation}
        \begin{aligned}
                \tilde{Z}' \ge 0 & \quad & \Rightarrow & \qquad & f(\phi(r)) \ge -\frac{1}{\gamma r^2} \\
                \tilde{Z}' < 0 & \quad & \Rightarrow & \qquad & f(\phi(r)) < -\frac{1}{\gamma r^2} 
        \end{aligned}
\end{equation}
For the existence of one or more Killing horizons it is required that the asymptotic region ($h \rightarrow 1$) is included in the first case where $h'$ is positive, too. The second case includes regions where $h'$ and $h$ are negative. If the Killing norm and its derivative have a mixed sign, no such statement is possible. Additionally, \eqref{eq:SRGKG} delivers more restrictions on the behaviour of $h$, $\phi$, $\phi'$ and $f$. After the introduction of the inverse radius ($\rho=1/r$, $\dot{\phi}:=\partial\phi/\partial\rho$) and with the requirement of asymptotic flatness ($0=f(\rho=0)$) the integration
\begin{equation}
        f(\phi)|_{\rho_h} - f(\phi)|_0 = \int\limits_0^{\rho_h}\!d\rho\frac{\rho^4}{(e)}\left(\dot{h}(\dot{\phi})^2 + h \dot{\phi} \ddot{\phi}\right),
\end{equation}
shows that $f(\rho_h)$ must be negative, if $\dot{\phi}$ and $\ddot{\phi}$ have a different sign (because $\dot{h}<0$ in the outermost region). These restrictions on the behaviour of the different occurring functions can be helpful while constructing solutions for the set of ODEs (\ref{eq:SRGHC}-\ref{eq:SRGKG}). 

In general it is very difficult to find an exact solution for a given $f(\phi)$. However, if a solution exists it will yield the scalar field $\phi$ as a function of the radius $r$. Thus, also $f$ can be expressed as a function of $r$. It is much more convenient to work with $f$ in this manner, because then the equations of motion can be reduced to a single ODE in one of the functions $h$, $(e)$, $\phi$ or $f$. Thereby one of these functions is assumed to be given and finally appears in different ways as coefficient function and as part of an eventual inhomogeneity in the ODE. For instance, one could take $h(r)$ as an input and solve the ensuing equation; clearly, in this manner the potential $f$ as a function of $\phi$ will not be an input but an outcome of the calculations. Such a procedure is similar to inverse scattering methods.

A useful possibility is to take $\phi$ to be given as function of the radius and try to calculate the remaining functions. Then \eqref{eq:SRGHC} establishes a simple connection between $(e)$ and $\phi$ and the combination of \eqref{eq:SRGSC} and \eqref{eq:SRGKG} leads to one linear second order ODE in $h$:
\begin{equation}
\label{eq:hDEn}
        h'' - h' r 2 \gamma (\phi')^2 - h \left(\frac{2}{r^2}+3\gamma (\phi')^2 + \gamma r \phi' \phi''\right) = -\frac{2(e)}{r^2}
\end{equation}
Thus, the whole problem is reduced to (a)~finding a ``plausible''
Ansatz for $\phi(r)$ and (b)~solving (\ref{eq:hDEn}) with ``physical'' boundary conditions, like $h(r\to\infty)=1$.

\subsubsection*{Example $\phi \propto 1/r$}
Now we specialize the discussion on the existence of Killing horizons to the choice of $\phi=\alpha/r$. This is a reasonable choice as the scalar field and its derivatives vanish sufficiently fast at infinity to allow for finite energy configurations. It is useful to employ the inverse radius $\rho=1/r$. Note that with $\phi(\rho)=\alpha \rho$ actually $f(\rho)$ essentially has the same form as $f(\phi)$.

The integration (cf. \eqref{eq:stat7}) of the Klein-Gordon equation \eqref{eq:SRGKG} from the asymptotic region $\rho=0$ to the Killing horizon $\rho=\rho_h$ simplifies to:
\begin{equation}
\label{eq:ShorKG}
        f(\phi)|_{\rho_h} - f(\phi)|_{\rho=0} = \alpha^2 \int\limits_0^{\rho_h}\! \frac{\rho^4}{(e)}\dot{h}\,d\rho
\end{equation}
Because we require $f(\phi(0))=f(0)=0$, and if $h$ has no local minima or maxima, the sign of $f(\rho_h)$ depends only on the sign of $\dot{h}$ in the region $\rho\in(0,\rho_h)$. We know that for an asymptotically flat metric the Killing norm must tend to $1$ at $\rho=0$. So between this point and the Killing horizon $h$ must be decreasing ($\dot{h}(\rho)<0$), which causes $f(\rho_h)$ to be negative. Moreover, the Klein-Gordon equation
\begin{equation}
        \alpha^2 \dot{h}(\rho) = \frac{(e)}{2 \rho^4} \dot{f}(\rho)
\end{equation}
shows that $\dot{h}$ and $\dot{f}$ have the same sign. Therefore, if $(e)$ is regular, points with $\dot{h}(\rho)=0$ imply $\dot{f}(\rho)=0$. According to this statement we can restrict $f$ to be negative for the whole interval $(0,\rho_h)$. Now the slicing condition \eqref{eq:slicing} 
\begin{equation}
\label{eq:ShorSC}
        \tilde{Z}' = -\rho \dot{h} + h = (e)\left[1 + \frac{1}{2\rho^2} f(\phi)\right]
\end{equation}
can be used to get another relation between $h$, $\dot{h}$ and $f$. From $\rho=0$ to the Killing horizon $\rho=\rho_h$ the function $h$ has to be positive and $\dot{h}$ negative. Thus the right hand side of \eqref{eq:ShorSC} has to be positive. With the above restriction of $f$ being negative there, the domain of the scalar potential in the interval $\rho \in (0,\rho_h)$ can be restricted to $-2\rho^2 < f(\rho) < 0$.

Now the existence of extremal Killing horizons and saddle points can be analyzed by inserting the choice of $\phi=\alpha \rho$ into \eqref{eq:hDEn} yielding
\begin{equation}
\label{eq:hDEn1}
        \ddot{h} + \dot{h} \frac{2}{\rho} \left[1+\frac{\alpha^2\rho^2}{2}\right]-h\frac{2}{\rho^2}\left[1+\frac{\alpha^2\rho^2}{4}\right] = - \frac{2(e)}{\rho^2}.
\end{equation}
To get a \textit{saddle point} $\ddot{h}$ and $\dot{h}$ have to vanish at some point $\rho=\rho_s$. Application of \eqref{eq:hDEn1} leads to the condition for $h(\rho_s)$ to be positive there. More precisely:
\begin{equation}
        h(\rho_{s}) = \frac{(e)}{1+\frac{\alpha^2\rho_{s}^2}{4}} \geq 0
\end{equation}
Obviously a saddle point that coincides with a Killing horizon ($h(\rho_s)=0$), is only possible for $\rho_s\to\infty$ or $(e)(\rho_s)=0$.

An \textit{extremal Killing horizon} occurs, where the Killing norm and its derivative vanish at the same value of $\rho=\rho_e$. Insertion into the ODE yields the condition
\begin{equation}
        \ddot{h}(\rho_e) = -\frac{2(e)}{\rho_e^2} \leq 0 \, .
\end{equation}
So the extremal Killing horizon has to be a maximum of $h$ and therefore must be in a region where the Killing norm is negative. Thus, for the present example an extremal Killing horizon cannot exist without a non-extremal outer one.

For a complete discussion of this solution \eqref{eq:hDEn1} has to be solved. It is easy to show that the homogeneous part can be transformed into a confluent hypergeometric differential equation with solutions in terms of $\chf(a,c;z)$.
Definitions and properties of the confluent hypergeometric function $\chf(a,c;z)$ are listed e.g. in \cite{slater:1960:bk}. With variation of constants the particular solution and therefore the general one can be determined up to an integration. 
\begin{figure}[htbp]
        \begin{center}
                \includegraphics[scale=0.95]{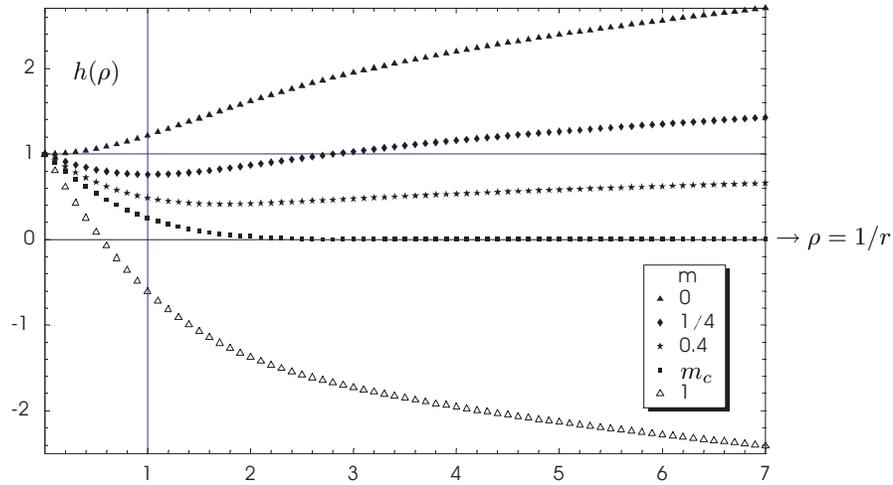}
                \caption{The numerical evaluation of $h(\rho)$ with $\phi=\rho=1/r$.}
                \label{fig:h_num}
        \end{center}
\end{figure}
\begin{figure}[htbp]
        \begin{center}
                \includegraphics[scale=0.95]{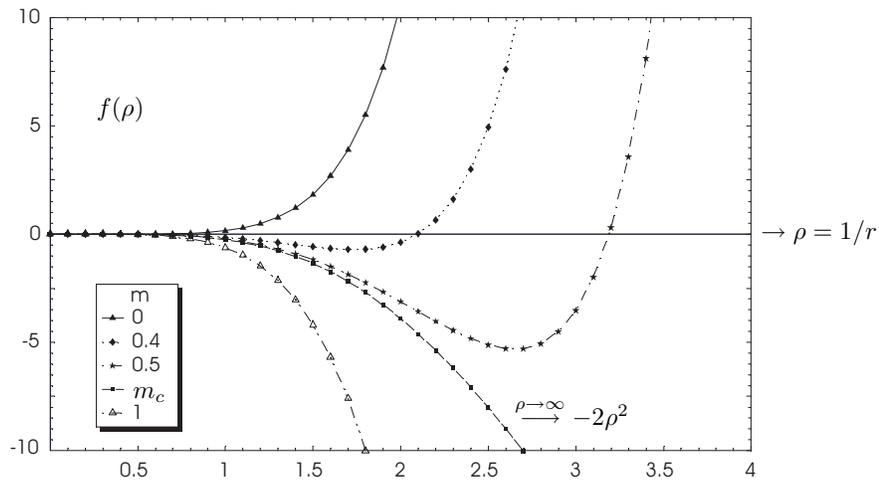}
                \caption{The corresponding numerical evaluation for $f(\rho)$.}
                \label{fig:f_num}
        \end{center}
\end{figure}

The perturbative and numerical evaluation of this solution as done in \cite{may04} shows\footnote{Note the slightly different conventions in \cite{may04}. Subsequent considerations are consistent with that conventions.} that there exists a critical mass above which Killing horizons are possible. Fig.~\ref{fig:h_num} exhibits the $m$-dependent behavior of the Killing norm (for $\al=1$). If $m<m_c$ no Killing horizons exist and the singularity at the origin becomes a naked one. The corresponding potential $f(\rho)=f(\phi)$ is depicted in fig.~\ref{fig:f_num}. If $m>m_c$ the potential is unbounded from below. Thus any black hole solution in the present case is likely to be unstable under non-static perturbations. For masses below that critical value a possibly stable solution but no Killing horizon exists, which is a counterexample to the cosmic censorship hypothesis. We hasten to add that in order to decide questions of stability of the solutions one has to give up the $2D$ formalism and study the full system of equations in $4D$.

\section{Generalizations}\label{se:3}

\newcommand{\tac}{T} 
\subsection{More general potentials}\label{sec:3.1}

Suppose instead of (\ref{vbh:matteraction}) a matter action
\begin{equation}
  \label{eq:sn1}
  L^{(m)} = \int\left[\frac 12 F(X)d\tac\wedge\ast d\tac + \eps f(X,\tac)\right]\,,
\end{equation}
where the scalar field now is denoted by $\tac$ to indicate its interpretation as tachyon in string theory. As geometric action (\ref{vbh:geometryaction}) will be taken with a {\em generic} potential of the form (\ref{vbh:calV}). The definitions
\eq{
I:=\exp{\int^X U(y)dy}\,,\quad w:=\int^XI(y)V(y)dy\,, 
}{defs}
are useful. The integration constants implicit in (\ref{defs}) are fixed according to the canonical mass definition derived in appendix \ref{app:mass}. Assuming staticity implies $d\tac=\tac'dX$; also all other fields solely depend on $X$, which later will be used as one of the coordinates.

The general classical solution can be obtained in analogy to sect.\ 3 of \cite{Grumiller:2002nm}: assuming that $X^+\neq 0$ in a given patch one defines a 1-form $Z:=e^+/X^+$ and proceeds to solve the EOM. The conservation equation reads
\begin{equation}
  \label{eq:sn10}
  d(I(X)X^+X^-+w(X)) + I(X)(X^+W^-+X^-W^+) = 0\,,
\end{equation}
with $W^\pm:=-F(\tac^\pm d\tac\mp e^\pm \tac^+\tac^-)\mp e^\pm f$ and $\tac^\pm:=\ast(d\tac\wedge e^\pm)= -X^\pm\tac'$. This yields
\begin{align}
W^+ &= F\left(X^+(\tac')^2dX + X^+X^-(\tac')^2 X^+ Z\right) - e^+ f\,,\\
W^- &= -FX^+X^-(\tac')^2 X^- Z + e^- f\,.
\end{align}
The only integration involved is $d\tilde{Z}=0\rightarrow\tilde{Z}=du$, where $Z=\tilde{I}(X)\tilde{Z}$. The result is ($A':=dA/dX$)
\begin{align}
e^- &= \frac{dX}{X^+} + X^-Z\,,\label{eq:sn3}\\
\eps &= e^+\wedge e^- = Z\wedge dX\,,\label{eq:sn4}\\
C &= \tilde{I}(X)X^+X^- + \tilde{w}(X)\,,\quad C = \rm const.\,,\label{eq:sn5}\\
\tilde{I}(X) &= \exp{\int^X\left(U(y)+F(y)(\tac')^2\right)dy}\,,\label{eq:sn6}\\
\tilde{w}(X)&= \int^X \tilde{I}(y)\left(V(y)+f(\tac,y)\right)dy \,,\label{eq:sn7}\\
0 &= \left(2FX^+X^-\tilde{I}\tac'\right)' - \tilde{I}\frac{\partial f}{\partial\tac}\,.\label{eq:sn8}
\end{align}
The last equation is the (generalized) Klein-Gordon equation. The line element reads
\begin{equation}
  \label{eq:sn2}
  ds^2=2dudr+K(X(r))du^2\,,\quad dr:=\tilde{I}(X)dX\,,\quad K(X(r))=2\tilde{I}(X) (C-\tilde{w}(X))\,.
\end{equation}
Note that in this construction $\tac'$ is assumed to be a known function of $X$ and $f(\tac,X)$ is then a derived quantity by means of (\ref{eq:sn8}). However, if the potential is known explicitly this method cannot be applied. Such an example will be studied in the next subsection.

There are the following obstructions to this construction
\blist
\item $X^+\neq 0$ in order to allow the definition of $Z$. If it vanishes and $X^-$ is nonvanishing everything can be repeated by $+\leftrightarrow -$. If both vanish in an open region a constant dilaton vacuum emerges. Such vacua are very simple, but their discussion will be omitted as the only solutions possible are (A)dS, Rindler and Minkowski.\footnote{\label{fn:cdv}If patched to a generic solution a constant dilaton vacuum induces a (anti-)selfdual matter flux along the patching line. This feature was found to be of relevance for ultrarelativistic boosts \cite{Balasin:2003cn}, in the classification of supersymmetric ground states of dilaton supergravity \cite{Bergamin:2003mh}, in the global discussion \cite{Grumiller:2003ad} of the ``kink'' solution in Kaluza-Klein reduced gravitational Chern-Simons theory \cite{Guralnik:2003we} and in the long time evaporation of the Schwarzschild BH \cite{Grumiller:2003hq}.} If both vanish at an isolated point something comparable to a bifurcation 2-sphere arises \cite{Klosch:1996fi,Klosch:1996qv}.
\item $\tilde{I}(X)$ is strictly monotonous in order to allow the redefinition from $X$ to $r$. To this end it is sufficient that $U(X)+F(X)(\tac')^2$ is strictly positive or negative, which is fulfilled e.g.~if $F$ and $U$ have the same sign everywhere.
\item For explicit calculations one needs $f$ to be a given function of $X$. Then one can deduce $\tac'$ by means of (\ref{eq:sn8}). Both $f$ and $\tac'$ are needed as a function of $X$ in order to perform the integrations in (\ref{eq:sn6}) and (\ref{eq:sn7}).
\elist

Note that further obstructions may occur: for instance, if $f=b\exp{(a\tac)}$, $a, b\in\mathbb{R}$, then (\ref{eq:sn8}) allows to express $\tilde{I}f$ in terms of $\tac'$ and thus the integration of the second term in (\ref{eq:sn7}) may be performed: $\int^X \tilde{I}f=2FX^+X^-\tilde{I}\tac'$. With the conservation law (\ref{eq:sn5}) this yields the obstruction $\tac' > - 1/(2F)$ in order to avoid singularities. Thus, the methods developed and explained in section~\ref{sec:SKH} by means of a pedagogical example readily generalize to generic dilaton gravity.

\subsection{Example: 2D type 0A with tachyon}

Recently in \cite{Douglas:2003up,Gukov:2003yp,Davis:2004xb} the low energy effective action for $2D$ type 0A string theory in the presence of RR fluxes has been studied (for simplicity an equal number $q$ of electric and magnetic D0 branes is assumed). The corresponding effective action up to second order in the tachyon in the second order formulation (translated to our notation) reads
\begin{multline}
  \label{eq:sn20}
  L=-\int d^2x\sqrt{-g}\Big[XR+\frac{(\nabla{X})^2}{X}-4\la^2X + \frac{\la^2q^2}{4\pi} \\
- X(\nabla\tac)^2 - \la^2X\tac^2 + \frac{\la^2q^2}{2\pi}\tac^2\Big]\,,
\end{multline}
with $\la^2=(26-D)/(12\al')=2/\al'$ and $\al'$ being the string coupling constant. Note especially the relative sign change in the potentials due to our conventions ($R$ is in our convention positive for negative curvature, in a desire to keep backward compatibility). The first four terms in (\ref{eq:sn20}) contain the ``geometric'' part, the rest the ``matter'' part. The translation into the first order form is straightforward,
\begin{equation}
  \label{eq:sn21}
  I(X)=\frac{1}{X}\,,\quad w(X)= - 2 \la^2 X  + \frac{\la^2q^2}{8\pi}\ln{X}\,,
\end{equation}
because $U=-1/X$ and $V=-2\la^2X+\la^2q^2/(8\pi)$. In the matter sector we have
\begin{equation}
  \label{eq:sn22}
  F(X)=-X\,,\quad f(\tac,X) = \frac12 \tac^2\left(\frac{\la^2q^2}{2\pi}-\la^2X\right)\,.
\end{equation}
Although the model (\ref{eq:sn20}) is not integrable, special cases can be treated in detail.

\paragraph{No tachyon} Without tachyon this model has been studied in \cite{Berkovits:2001tg}. The geometric part is essentially the Witten BH \cite{Mandal:Elitzur:Witten} plus an extra term from the RR fluxes. It can be treated classically, semi-classically and at the quantum level by standard methods available for dilaton gravity \cite{Grumiller:2002nm}. 

\paragraph{Constant tachyon} Solutions with constant tachyon have been investigated in \cite{Thompson:2003fz}. Analyzing the equations of motion it can be shown easily that a constant tachyon implies a constant dilaton and vice versa. Thus, solutions with constant tachyon are constant dilaton vacua, implying $X^\pm=0$. They have been encountered recently in various instances, cf.~footnote \ref{fn:cdv}. Geometry turns out to be either Minkowski, Rindler or (A)dS spacetime, as curvature is given by
\begin{equation}
  \label{eq:cdv13}
  R = - 2 V'(X) - \frac{\partial}{\partial X}f(\tac,X) = \rm const.
\end{equation}
The conservation equation (\ref{eq:sn10}), which simplifies to $V+f=0$, together with the ``Klein-Gordon'' equation (\ref{eq:sn8}), which simplifies to $\partial f/\partial T=0$, establishes relations to be fulfilled by the dilaton and the tachyon. It turns out that besides $T=0, X=q^2/(16\pi)$ (in agreement with (3.12) of \cite{Thompson:2003fz}) no regular solution exists. However, one can easily generalize the discussion to generic $f=f_0(T)-Xf_1(T)$. The equations of motion yield a constraint to be fulfilled by the tachyon potentials ($f'_i:=df_i/dT$): 
\begin{equation}
  \label{eq:cdv17}
  \frac{f_0'}{f_1'} = \frac{f_0+\la^2q^2/(8\pi)}{f_1+2\la^2}
\end{equation}
It is emphasized that \eqref{eq:cdv17} need not be fulfilled for all values of $T$, but just at isolated points $T=T_0$. Additionally the value of the dilaton $X$ at these points can be calculated:
\begin{equation}
	\label{eq:cdv17.1}
	X=\frac{f_0(T_0)+\lambda^2q^2/(8\pi)}{f_1(T_0)+2\lambda^2}
\end{equation}
It is not guaranteed that every solution of \eqref{eq:cdv17} produces a positive dilaton. However, if $f_0>-\lambda^2q^2/(8\pi)$ and $f_1>-2\lambda^2$ for all values of $T$ then positivity of $X$ is ensured for all solutions with constant tachyon. 

For the choice $f_0=\la^2q^2/(8\pi) (\cosh{(2T)}-1)$ advocated in \cite{Douglas:2003up} eq.~(\ref{eq:cdv17}) simplifies to
\begin{equation}
  \label{eq:cdv23}
  2\tanh{(2\tac)} = \frac{f_1'}{2\la^2 + f_1}
\end{equation}
Perturbatively, $f_1 = \tac^2 \la^2/2 - \tac^4 c_1/4$ with some unknown constant $c_1$; to leading order this yields in addition to $T=0$ constant dilaton vacua with
\begin{equation}
  \label{eq:cdv20}
  T = \pm \sqrt{\frac{84}{125-12 c_1/\la^2}} \approx \pm 0.6 + \mathcal{O}(c_1/\la^2)
\end{equation}
However, the existence of these solutions is an artifact of perturbation theory; only for very large negative $c_1$ a solution close to $T=0$ may appear. Thus, the eventual appearance of constant dilaton vacua with $T\neq 0$ is a nonperturbative effect. Note that for small $T$ the l.h.s. in (\ref{eq:cdv23}) is larger than the r.h.s.---thus, if at least the asymptotic behavior of $f_1(T\gg 1)$ can be extracted by any means this guarantees the existence of at least one such vacuum provided that the asymptotics implies that the l.h.s. is now smaller than the r.h.s.: by continuity at least one zero at finite $T$ has to exist.
For instance, if $\lim_{T\to\infty} f_1= a e^{bT}$ with positive $a$ and $b>2$ by the previous arguments we know that at least one non-trivial solution with constant tachyon and dilaton must exist.

\paragraph{Static tachyon} Asymptotic ($X\to\infty$) solutions with static tachyon have been presented in appendix B of \cite{Davis:2004xb}; it has been found that the tachyon vanishes asymptotically. 

Now we apply the previous results of this work. Self-interactions of the tachyon do exist and the Jacobi matrix has maximal rank. Regarding positivity obstructions it should be noted that $F$ is negative. Thus, $\sqrt{-g}=(e)$ is positive and monotonous. However, neither $f$ nor $V$ have a definite sign. Consequently, the existence of Killing horizons cannot be ruled out by any of the criteria discussed in section \ref{se:2}.

Therefore, let us solve the equations of motion on a Killing horizon, i.e.~$X^+X^-=0$. This can serve as a basis of a perturbative analysis around the Killing horizon in analogy to \cite{Filippov:2002sp}. Then the conservation equation (\ref{eq:sn10}) simplifies to the one encountered in the discussion of constant dilaton vacua,
\eq{
V+f=0\quad\rightarrow\quad 2\tac^2 =  \frac{8\xi-1}{1-\xi}\,,\quad \xi:=\frac{2\pi X}{q^2}\,,
}{eq:sn23} 
where all quantities have to be evaluated at the Killing horizon. Only those values of the ADM mass (cf.~appendix \ref{app:mass}) which lead to a global behavior of the tachyon as a function of the dilaton capable to satisfy (\ref{eq:sn23}) can create a static solution with a Killing horizon. 

Obviously, in the asymptotic region $X\gg 1$ no Killing horizon may emerge for a real tachyon. To be more precise, $X$ has to be smaller than $q^2/(2\pi)$. Thus, unless a large $q$ expansion is invoked, the existence of eventual Killing horizons is a nonperturbative effect. In that case higher order terms in $T$ in the action (\ref{eq:sn20}) which have been neglected so far can play a crucial role.

It could be interesting to perform steps analogous to the ones in section \ref{sec:horSRG}, i.e.~to find a solution which is asymptotically flat, has an asymptotically vanishing tachyon and a Killing horizon. One can take the leading order tachyon potential as in (\ref{eq:sn20}) allowing for an unknown addition containing higher powers in the tachyon. Then, by cleverly\footnote{For instance, $T$ should behave asymptotically like in appendix B of \cite{Davis:2004xb}.} designing $T(X)$ one can, at least in principle, solve the full system of equations of motion and afterwards read off $f$ as a function of $X$. Finally, this has to be translated into the form $f=f_0(T)-Xf_1(T)$, such that to leading order in $T^2$ eq.~(\ref{eq:sn22}) is reproduced.

\subsection{Concluding remarks}

Exploiting the technical advantages inherent to the first order formulation of $2D$ dilaton gravity we have studied in detail static solutions of generic dilaton gravity coupled to self-interacting scalar matter. As explained in more detail in the introduction our main motivation was to find under which conditions static BHs with scalar hair may exit. The rank of the Jacobi matrix of the underlying dynamical system played a crucial role.
With some (pathological) exceptions occurring in the rank 2 case, only the rank 3 case may lead to regular nontrivial Killing horizons and thus only one constant of motion may exist for hairy black hole solutions. It can be interpreted as the mass of the spacetime, a notion which is not completely trivial and therefore studied in detail in the appendix. An example relevant to spherically symmetric general relativity has been provided explicitly. 

We were able to generalize our results on the (non-)existence of hairy black hole solutions (immersed in flat spacetime) to generic dilaton gravity \eqref{vbh:geometryaction}, \eqref{vbh:calV} with matter action \eqref{eq:sn1}. As a particular example $2D$ type 0A string theory has been discussed from the perspective developed in our work. It has been found for this model that, regardless of the specific form of the tachyon potential, no Killing horizon may emerge in the asymptotic region (unless a large $q$ expansion is invoked, where $q$ is the number of electric and magnetic D0 branes). A method has been suggested which works nonperturbatively; however, it involves the ''clever design'' of one arbitrary function.

Straightforward applications not discussed so far are spherically symmetric scalar tensor theories (some examples can be found in \cite{Salgado:2003ub}). Their relation to dilaton gravity with matter has been studied in \cite{Grumiller:2000wt}.

In addition to the generalizations discussed above one can add further matter fields or gauge fields. The latter essentially deform the geometric potentials, as they will not contribute to physical propagating degrees of freedom\footnote{This is true for intrinsically two dimensional gauge fields. It remains true in the context of certain dimensional reductions, e.g. spherically reduced $U(1)$ gauge fields, but it does not hold in general. Probably the best known counter example is the Bartnik-McKinnon soliton \cite{Bartnik:1988am} (for a review cf.~ref.~\cite{Volkov:1998cc}).}; thus, this generalization is already covered by our current discussion. The former will enlarge the physical phase space and thus a more complicated analysis than the one presented in this work may be necessary; in particular, it could be of interest to study the behavior of the rank of the Jacobi matrix of the corresponding dynamical system and its impact on the (non-)existence of Killing horizons.

Finally, it should be pointed out that the analysis was purely twodimensional. Thus, examples which are relevant to general relativity cannot deal with stability questions regarding non-spherically symmetric perturbations. It could be argued on general grounds---invoking the no-hair conjecture---that the hairy black holes implied by the present work are unstable against such perturbations. However, such an analysis necessarily has to be a fourdimensional one.

\section*{Acknowledgements}
 
This work has been supported by project P-14650-TPH and J-2330-N08 of the Austrian Science Foundation (FWF). We thank D.~Maison, D.~M.~Thompson and L.~A.~Pando Zayas for correspondence and A.~Filippov, W.~Kummer and D.~Vassilevich for discussion. One of the authors (DG) is grateful to M.~Katanaev for drawing his attention to static solutions and pointing out ref.\ \cite{Fisher:1948yn}. Part of this work was performed in the hospitable atmosphere of the Erwin Schr{\"o}dinger Institut during the workshop ``Gravity in two dimensions'' in fall 2003. 

\begin{appendix}

\section{Canonical mass definition}\label{app:mass}

The question of how to define ``the'' mass in theories of gravity is notoriously cumbersome. A nice clarification for $D=4$ is contained in ref.\ \cite{Faddeev:1982id}. The main conceptual point is that any mass definition is meaningless without specifying (a)~the ground state spacetime with respect to which mass is being measured and (b)~the physical scale in which mass units are being measured. Especially the first point is emphasized here.

In addition to being relevant on its own, a proper mass definition is a pivotal ingredient for any thermodynamical study of BHs. Obviously, any mass-to-temperature relation is meaningless without defining the former (and the latter). For a large class of $2D$ dilaton gravities these issues have been resolved in ref.\ \cite{Liebl:1997ti}. One of the key ingredients is the existence of a conserved quantity \cite{Banks:Frolov:Mann} which has a deeper explanation in the context of first order gravity \cite{Grosse:1992vc} and PSMs \cite{Schaller:1994es}. It establishes the necessary prerequisite for all mass definitions, but by itself it does not yet constitute one. Ground state and scale still have to be defined.

This section provides a canonical prescription to obtain ``the'' mass for generic $2D$ dilaton gravity (\ref{vbh:geometryaction}), (\ref{vbh:calV}). A list of examples is provided, including $2D$ type 0A strings where apparently mutually contradicting results for the ADM mass exist in the literature \cite{Gukov:2003yp,Davis:2004xb,Danielsson:2004xf}. It will be shown in the fourth example that actually all of them are correct.

\paragraph{Zeroth step: definitions} It is useful to define the functions
\begin{equation}
  \label{eq:2d1}
  \bar{I}(X):=\exp{\int^X U(y)dy}\,,\quad \bar{w}(X):=\int^y\bar{I}(y)V(y)dy\,,
\end{equation}
where the integration constants are fixed by requiring that no contribution from the lower boundary arises. E.g.~for $U=-1/X$ the quantity $\bar{I}=1/X$ does not acquire any multiplicative factor. Note that shifts in $\bar{w}$ correspond to shifts in the ground state energy while rescalings in $\bar{I}$ correspond to a change of mass units. Below it will be discussed how to fix these ambiguities.

\paragraph{First step: identify ground state} Once a ground state geometry is identified the canonical prescription is to set its value of the Casimir function \cite{Schaller:1994es}
\begin{equation}
  \label{eq:2d2}
  \bar{C}:=\bar{I}X^+X^- + \tilde{w}
\end{equation}
to zero by shifting $\bar{w}\to\tilde{w}=\bar{w}+\rm const.$ The tricky part, of course, is to identify the ground state. Here is a guideline how to proceed: for models with Minkowskian, Rindler or (A)dS solution in the spectrum it is natural to take this spacetime as ground state. For models which allow a SUGRA extension, i.e.~$w(X)\leq 0, \forall X$, no such shift is needed as the definition in (\ref{eq:2d2}) is already the correct one (up to rescaling to be addressed in the second step). If 1/2 BPS states exist they will be the unambiguous ground state \cite{Bergamin:2003mh}. In generic non-SUGRA theories which allow no SUGRA extension it need not be possible to define a ``natural'' ground state. In that case one has to take any solution, declare it as ground state by definition and calculate masses with respect to it. For models which are asymptotically Minkowski but do not have a Minkowskian ground state (like $2D$ type 0A with nonvanishing RR fluxes, for instance) a natural choice for a ground state could be a solution where the Killing norm depends only on the ``charges'' of the model. Sometimes two or more solutions might be candidates for the gound state. Especially in such cases it is important to specify it explicitly.

\paragraph{Second step: identify asymptotic region(s)} These are regions where $|\bar{w}(X)|\gg 1$ and they are located typically at $X\to 0$ or $X\to\infty$. For a given model exactly one, several or no such regions might exist. In the latter case no mass can be defined in this manner. In the second case one has a separate mass definition for each asymptotic region, in complete analogy to the first case which will be discussed now. It is proposed to take the Killing norm of the ground state $-2\bar{I}\tilde{w}$ and to expand it in the asymptotic region in powers of $\bar{r}$, where $d\bar{r}=\bar{I}dX$ (as usual it will be assumed that $\bar{I}$ is strictly monotonic; then, this redefinition is well-defined). Actually, only the leading order term is needed, so the next-to-leading order may contain non-analytic expressions. For definiteness it will be supposed that the asymptotic region lies at $\bar{r}\to\infty$.
\begin{equation}
  \label{eq:2d3}
  -2\bar{I}\tilde{w} = \la^2 \bar{r}^\al \left(1 + f(\bar{r}))\right)\,,\quad\al,\la\in\mathbb{R}\,,\quad\lim_{\bar{r}\to\infty}f(\bar{r})=0\,.
\end{equation}
A rescaling prescription which fixes the physical scale uniquely is the requirement $\la=1$, i.e.~$\bar{I}\to I=\la^{-1}\bar{I}$ which implies $\tilde{w}\to w=\la^{-1}\tilde{w}$. This somewhat arbitrary prescription yields the mass in units of the asymptotic Killing time for asymptotically flat spacetimes (i.e.~``the'' ADM mass), proportional to the asymptotic acceleration for asymptotically Rindler spacetimes and in terms of the absolute value of the asymptotic curvature for asymptotically (A)dS. If the leading order term is not a monomial, one still defines a certain scale in the same manner---actually, it does not matter very much how it is defined but only {\em that} it is defined! The physical meaning of the rescaling can be demonstrated most easily by comparing the original line element \cite{Klosch:1996fi},
\begin{equation}
  \label{eq:2d10}
  ds^2 = 2d\bar{u}d\bar{r}+(-2\bar{I}\bar{w})\left(1+\frac{\bar{C}}{\bar{w}}\right)d\bar{u}^2\,,
\end{equation}
with the rescaled one
\begin{equation}
  \label{eq:2d11}
  ds^2 = 2dudr+(-2Iw)\left(1+\frac{C}{w}\right)du^2\,.
\end{equation}
They coincide if $r=\bar{r}\la^{-1}$, $u=\bar{u}\la$ and $C=\bar{C}\la^{-1}$. Thus, the units of ``radius'', one over ``advanced time'' and ``mass'' have been rescaled by the same factor. 

\paragraph{Third step: a unique mass definition} Having fixed the shift and scale ambiguity as discussed above the mass is given uniquely by the negative Casimir function
\begin{equation}
  \label{eq:2d4}
  M:= - C = - I(X)X^+X^- - w(X) = -\bar{C}\la^{-1}\,,
\end{equation}
and thus the line element reads 
\begin{equation}
  \label{eq:2d5}
  ds^2 = 2dudr + (-2Iw)\left(1-\frac{2M}{(-2w)}\right)du^2\,,\quad I(X)dX=dr\,.
\end{equation}
In summary, as compared to the original definition (\ref{eq:2d2}) by applying the proposed procedure $\bar{C}$ has been shifted by $\tilde{w}-\bar{w}$ and then rescaled by $\la^{-1}$ from (\ref{eq:2d3}). 

\paragraph{Example 1: recovering ADM} The most important check is whether the construction proposed is able to recover the ADM mass \cite{Arnowitt:1962} in the context of $2D$ dilaton gravity \cite{dilaton,Louko:1995tv,Kummer:1997si,Liebl:1997ti}. This can be proven most easily by starting from (5.10) of \cite{Grumiller:2002nm} and recalling that the square root of the Killing norm in (\ref{eq:2d5}) for $-2Iw=1$ is given by $\sqrt{1+M/w}= 1 - M/(2w) + \mathcal{O}(M^2/w^2)$. With $\partial_r X=I^{-1}(X)$ the result is $M_{ADM} = M$, with $M$ from (\ref{eq:2d4}). Note however, that an additional rescaling may be needed if overall factors in front of the action have been dropped (cf.~the brief discussion for the Schwarzschild BH between (5.11) and (5.12) of \cite{Grumiller:2002nm}).

\paragraph{Example 2: the $\boldsymbol{a-b}$ family} This useful family \cite{Katanaev:1997ni} encompasses many prominent models. The ensuing mass definition was discussed in \cite{Liebl:1997ti}. It is shown now that the results can be reproduced easily with the prescription above. Zeroth step: for this family the relevant functions read $\bar{I}=X^{-a}$ and $\bar{w}=-BX^{b+1}/(2(b+1))$ (cf.~(3.65) of \cite{Grumiller:2002nm}). First step: no shift is needed as the functions defined in (\ref{eq:2d1}) already provide the correct ground state. Second step: with (3.65) of \cite{Grumiller:2002nm} the monomial $-2\bar{I}\bar{w}=B/(b+1) \cdot ((1-a)\bar{r})^{(b-a+1)/(1-a)}$ implies $\la=\sqrt{B/(b+1)}(1-a)^{(b+1-a)/(2(1-a))}$ (assuming $a\neq 1$, $b\neq -1$ and $B\neq 0$). Thus, the mass $M$ in terms of the original value of the Casimir $\bar{C}$ is given by
\begin{equation}
  \label{eq:2d6}
  M=-C=-\bar{C}\la^{-1}=-\bar{C}\sqrt{\frac{b+1}{B}}(1-a)^{(a-b-1)/(2(1-a))}\,.
\end{equation}
For Minkowski ground state models $a=b+1$ this coincides with (5.11) of \cite{Grumiller:2002nm}. Up to a numerical factor which is due to different conventions (\ref{eq:2d6}) agrees with (40) of \cite{Liebl:1997ti}. For the Rindler case $b=0$ the scale $\la$ turns out to be twice the asymptotic Rindler acceleration, because $ds^2=2dudr+\bar{r}du^2=2dudr+(\la r)du^2$ is diffeomorphic to $ds^2=(R\la/2)^2dT^2-dR^2$. For (A)dS $b=1-a$ and thus $\la^2$ is twice the absolute value of asymptotic curvature because $ds^2=2dudr+\bar{r}^2du^2=2dudr+(\la r)^2du^2$. 
Let us finally address the exceptional cases omitted so far: for $B=0$ the model is flat on all solutions and hence no meaningful mass definition exists. For $b=-1$ the quantity $w\propto\ln{X}$ does not lead to monomial behavior as in (\ref{eq:2d3}), apart from $a=0$ (conformally transformed CGHS), but nonetheless a mass scale can be defined. For $a=1$ and $b=0$ (CGHS) no problem arises as the theory belongs to the MGS class, while for $a=1$ and $b\neq 0$ similar remarks apply as for $b=-1$ and $a\neq 0$.  

\paragraph{Example 3: SUGRA with 1/2 BPS} Suppose $\bar{w}\leq 0$ for all $X$. Then, the SUGRA restrictions are fulfilled, i.e.~a prepotential exists \cite{Ertl:2000si,Bergamin:2002ju}, and $\bar{w}=\tilde{w}$, thus fixing the ground state. The BPS condition reads \cite{Bergamin:2003mh} $\bar{C}=C=0$. The scale can be fixed in analogy to previously discussed cases. In this case a positive mass theorem exists \cite{Park:1993sd} in analogy to Witten's proof \cite{Witten:1981mf}.

\paragraph{Example 4: $2D$ type 0A} For the matrix model description of $2D$ type 0A/0B string theory we refer to \cite{Takayanagi:2003sm,Douglas:2003up} (for an extensive review on Liouville theory and its relation to matrix models and strings in $2D$ cf.~\cite{Nakayama:2004vk}). We will focus on the target space description (2.9) of \cite{Douglas:2003up}. In order to define the geometric mass the tachyon can be set to zero. The relevant functions read $\bar{I}=1/X$ and $\bar{w}=-2\la^2 X + \la^2q^2/(8\pi) \cdot \ln(X)$. The tricky issue is that there are two natural candidates: on the one hand, the model is asymptotically flat, on the other hand a BPS state exists. Actually, there is a well-known precedent to this: the Reissner-Nordstr\"om BH with Killing norm $1-2M/r+Q^2/r^2$ allows also for two mass definitions: the ADM mass $M_{ADM}=M$ or the ``BPS mass'' (which vanishes for extremal BHs of a given charge) $M_{BPS}=M-|Q|$. Naturally, the ``BPS mass'' is lower than the ADM mass. Obviously for each value of $Q$ there will be a different shift. It is okay to take either of them, but one has to specify {\em which}. In the same manner for the model under consideration the ADM mass is given by $M_{ADM}=-\bar{C}/(2\la)$ in agreement with \cite{Davis:2004xb,Danielsson:2004xf} (up to notations). The condition of extremality reads $M_{ADM}^{(BPS)}=(1-\ln{(X_0)})q^2/(16\pi)$ where $X_0=q^2/(16\pi)$ is the value of the dilaton at the Killing horizon. The ADM mass of a BPS state is positive (negative) for $X_0>e$ ($X_0<e$). The BPS mass is shifted such that it always vanishes for extremal BHs, i.e.
\begin{equation}
  \label{eq:2d12}
  M_{BPS}:=M_{ADM}-M_{ADM}^{(BPS)} = M_{ADM} - \frac{q^2}{16\pi}\left(1-\ln{\frac{q^2}{16\pi}}\right)\,.
\end{equation}
In the large $q$ expansion this mass definition coincides with (2.38) of ref.\ \cite{Gukov:2003yp} up to notational differences and a sign. The sign can be reversed by taking $M_{BPS}$ as input and expressing the ADM mass in the large $q$ limit. In this sense, despite of their differences the papers \cite{Gukov:2003yp,Davis:2004xb,Danielsson:2004xf} produce all the correct ADM mass. Note that SUGRA demands $q^2\leq 16\pi e$ \cite{Bergamin:2003mh}, a bound that has been found first in \cite{Davis:2004xb}. Thus, for solutions consistent with SUGRA the term in the bracket in (\ref{eq:2d12}) is nonnegative and therefore, like in the Reissner-Nordstr{\"o}m case above, the ``BPS mass'' is always smaller than the ADM mass. Clearly a large $q$ expansion is not applicable in that case.

\end{appendix}


\providecommand{\href}[2]{#2}

\end{document}